\documentclass{article}
\usepackage[utf8]{inputenc}
\usepackage{amsmath}
\usepackage{amsfonts}
\usepackage{mathrsfs}
\newtheorem{definition}{Definition}
\newtheorem{proposition}{Proposition}
\newtheorem{theorem}{Theorem}

\newtheorem{lemma}{Lemma}
\usepackage[margin=1in]{geometry}
\usepackage{multirow}
\usepackage{qcircuit}
\usepackage{graphicx}
\usepackage{hyperref}
\usepackage{authblk}

\title{S-FABLE and LS-FABLE: Fast approximate block-encoding algorithms for unstructured sparse matrices}
\author[1]{Parker Kuklinski}
\author[1]{Benjamin Rempfer}
\affil[1]{MIT Lincoln Laboratory, Lexington, MA, USA}
\date{October 17, 2023}

\begin{document}

\maketitle

\begin{abstract}
The Fast Approximate BLock-Encoding algorithm (FABLE) is a technique to block-encode arbitrary $N\times N$ dense matrices into quantum circuits using at most $O(N^2)$ one and two-qubit gates and $\mathcal{O}(N^2\log{N})$ classical operations. The method nontrivially transforms a matrix $A$ into a collection of angles to be implemented in a sequence of $y$-rotation gates within the block-encoding circuit. If an angle falls below a threshold value, its corresponding rotation gate may be eliminated without significantly impacting the accuracy of the encoding. Ideally many of these rotation gates may be eliminated at little cost to the accuracy of the block-encoding such that quantum resources are minimized. In this paper we describe two modifications of FABLE to efficiently encode sparse matrices; in the first method termed Sparse-FABLE (S-FABLE), for a generic unstructured sparse matrix $A$ we use FABLE to block encode the Hadamard-conjugated matrix $H^{\otimes n}AH^{\otimes n}$ (computed with $\mathcal{O}(N^2\log N)$ classical operations) and conjugate the resulting circuit with $n$ extra Hadamard gates on each side to reclaim a block-approximation to $A$. We demonstrate that the FABLE circuits corresponding to block-encoding $H^{\otimes n}AH^{\otimes n}$ significantly compress and that overall scaling is empirically favorable (i.e. using S-FABLE to block-encode a sparse matrix with $\mathcal{O}(N)$ nonzero entries requires approximately $\mathcal{O}(N)$ rotation gates and $\mathcal{O}(N\log N)$ CNOT gates). In the second method called `Lazy' Sparse-FABLE (LS-FABLE), we eliminate the quadratic classical overhead altogether by directly implementing scaled entries of the sparse matrix $A$ in the rotation gates of the S-FABLE oracle. This leads to a slightly less accurate block-encoding than S-FABLE, while still demonstrating favorable scaling to FABLE similar to that found in S-FABLE. These sparse encoding methods work best on unstructured data and lose their efficacy when structure or symmetry is introduced.
\end{abstract}

\section{Introduction}

Block-encoding classical data into a quantum circuit has become a central component of modern quantum algorithms \cite{clader22}. The technique is used in algorithms such as quantum phase estimation \cite{morisaki23}, quantum singular value transformations \cite{lloyd21}, quantum differential equations solvers \cite{fang23}, and other linear systems solvers \cite{wan21}. Block-encoding entails embedding a matrix $A$ (not necessarily unitary or Hermitian) into a larger unitary matrix $U$ which can be constructed via a quantum circuit. After preparing an initial state and performing the circuit, a successful measurement on the ancilla qubits returns the matrix $A$ applied to the initial state vector as a quantum state, which can then be used for the purpose of a particular algorithm.

While many quantum algorithms assume the existence of certain block-encoded matrices, the task of constructing a block-encoding is nontrivial. Algorithms exist for efficient block-encoding of matrices with well-defined structure, such as tridiagonal matrices \cite{sunderhauf23}, matrices corresponding to graphs such as binary trees \cite{camps22}, heierarchical matrices \cite{nguyen22}, and matrices corresponding to pseudodifferential operators \cite{li23}. Comparatively few methods exist for block-encoding dense matrices; see \cite{clader22} for a review of dense block-encodings for the express purpose of implementing QRAM based algorithms.

One such method for block-encoding dense matrices is the Fast Approximate BLock-Encoding algorithm (FABLE) introduced by Camps and Van Beeumen \cite{fable22}, adapted from an earlier method by Mottonen et. al. \cite{mottonen04}. The FABLE algorithm implements an oracle by converting entries of an $N\times N$ matrix $A=(a_{ij})$ into rotation angles via an inverse cosine, i.e. $\theta_{ij}=\cos ^{-1}a_{ij}$. Rather than naively implementing these rotations as multi-qubit controls, the FABLE oracle alternates single-qubit rotation gates with CNOT gates whose controls follow a Gray code pattern, leading to $\mathcal{O}(N^2)$ one and two qubit gate complexity. However, instead of directly using the derived angles $\theta _{ij}$ this oracle uses a transformed set of angles $\hat{\theta}$ defined by the linear transformation $NH^{\otimes 2n}P_G\text{vec}(\hat{\theta})=\text{vec} (\theta )$ where $P_G$ is a Gray code permutation and $H$ is the $2\times 2$ Hadamard matrix. Computing the transformed angles $\hat{\theta}$ requires $\mathcal{O}(N^2\log{N})$ classical operations.

While in the worst case the FABLE oracle has $2N^2$ gates, in other situations we can significantly reduce this count at little cost to the accuracy of the block-encoding. If any of the transformed angles have absolute value below a threshold quantity $\delta$, then we can eliminate those rotation gates at a cost on the $L^2$ norm of the approximation bounded by $N^3\delta$ (this bound is orders of magnitude larger than the typical error). Furthermore, since all of the CNOT gates in the oracle have the same target qubit, sequences of these CNOT gates can commute and cancel, leading to further compression on the gate count of the oracle.

An open question left in \cite{fable22} asks about the characterization of matrices which have highly compressible FABLE circuits. Several times in the paper it is mentioned that matrices which are sparse in the Walsh-Hadamard domain appear to perform well under this block-encoding, i.e. matrices such that $H^{\otimes n}AH^{\otimes n}$ is sparse. However, as discussed in \cite{fable22} the connection between problem domain and matrices sparse in the Walsh-Hadamard domain is unclear; several matrices corresponding to well known quantum algorithms seem to require a large fraction of the FABLE oracle to block-encode well.

In this paper we pose a different question; how can we efficiently transform application-relevant matrices into Walsh-Hadamard sparse matrices which compress well with FABLE? We begin by noting that generic sparse matrices are ubiquitous in quantum computing algorithms \cite{bellante22, legall17, berry07, wang17}. While some structured sparse matrices such as those previously mentioned admit efficient block-encodings, deviations from this structure do not bode well for these kinds of bespoke encodings. Furthermore, sparse matrices in general do not have well compressed FABLE circuits. Instead, one would hope in the general case that the small number of classical resources required to describe a sparse matrix may translate into a commensurately low requirement on the quantum resources to encode that same matrix. 

A modification of the FABLE algorithm, which we call Sparse-FABLE (S-FABLE), demonstrates this intuition to be correct. If $A$ is a sparse matrix, then the matrix $H^{\otimes n}AH^{\otimes n}$ is sparse in the Walsh-Hadamard domain (i.e. $H^{\otimes n} (H^{\otimes n}AH^{\otimes n} )H^{\otimes n} =A$ is sparse) and compresses well in the FABLE framework. Thus, to block encode a sparse matrix $A$ with S-FABLE we block-encode $H^{\otimes n}AH^{\otimes n}$ with a standard FABLE circuit and then conjugate the circuit by $n$ extra Hadamard gates both sides. At the cost of classically computing $H^{\otimes n}AH^{\otimes n}$ (which takes $\mathcal{O}(N^2\log N)$ operations) and $2n$ additional Hadamard gates, we can drastically compress the size of the FABLE oracle, potentially at an exponential advantage.

While the empirical contraction of quantum resources used by S-FABLE over FABLE is promising, both methods still require a large amount of classical pre-processing to compute the angles used in the rotation gates of the oracles. To this end, we introduce another method called `Lazy' Sparse-FABLE (LS-FABLE), a first order approximation of S-FABLE which necessitates only as many classical computations as there are nonzero entries in the sparse matrix being encoded, and uses that same $\mathcal{O}(N)$ number of rotation gates in its oracle. This appeals to the intuition that the resources used to block-encode a matrix (both quantum and classical) should approximately scale with the resources needed simply to describe the matrix. We find the cost of the LS-FABLE approximation is close to a constant factor decrease in accuracy over S-FABLE, however the beneficial scaling properties of S-FABLE are retained.

The improvement of these new sparse modifications of FABLE over the original algorithm are not universal, however. Both S-FABLE and LS-FABLE succeed in block-encoding unstructured random matrices with entries uniformly randomly distributed from $[-1,1]$; any deviation from this and these new algorithms lose their advantage. For example, S-FABLE and LS-FABLE have difficulty block-encoding graph adjacency matrices with nonzero entries equal to one, and even more disappointingly unstructured nonnegative sparse matrices, requiring nearly all of their rotation gates to block-encode these classes of matrices to moderate accuracy. These methods also struggle in the presence of symmetry, as in matrices arising from physical problems like Heisenberg model Hamiltonians or simple Laplacians; in these cases FABLE strongly outperforms S-FABLE and LS-FABLE. 

The remainder of this paper is structured as follows; section \hyperref[FABLE]{(2)} details the original FABLE algorithm and important results from \cite{fable22}, section \hyperref[SFABLE]{(3)} documents the new S-FABLE algorithm for unstructured sparse matrices, section \hyperref[LSFABLE]{(4)} introduces LS-FABLE, an approximation of S-FABLE requiring only $\mathcal{O}(N)$ classical operations, and section \hyperref[DATA]{(5)} illustrates numerical results of the implementation for a variety of classes of matrices.

\section{FABLE: Fast Approximate Block Encoding Algorithm}\label{FABLE}

Let $A=(a_{ij})$ be a real $N\times N$ matrix with $N=2^n$ and entries satisfying $a_{ij}\in [-1,1]$. We formally define the concept of a block-encoding, or embedding a matrix into a larger unitary operator:
\begin{definition}
Let $a,n\in\mathbb{N}$ and $m=a+n$. We say that an $m$-qubit unitary operator $U$ is a $(\alpha ,a,\epsilon )$-\emph{block-encoding} of an n-qubit operator $A$ (not necessarily unitary) if $\lVert A-\alpha\tilde{A}\rVert _2<\epsilon$ where
$$\tilde{A}=\left(\langle 0|^{\otimes m}\otimes I_n\right) U\left( |0\rangle ^{\otimes m}\otimes I_n\right)$$
\end{definition}
Visually we can depict this unitary as
$$U=\begin{pmatrix} \tilde{A} & \cdot \\ \cdot & \cdot\end{pmatrix}$$
where the top left corner contains a scaled approximation of the operator to be block-encoded while the remainder of the matrix contains `junk'. If we operate $U$ on a state $|0\rangle ^{\otimes m}\otimes |\psi\rangle$ and measure the ancilla register to be in $|0\rangle ^{\otimes m}$ (which occurs with probability $\lVert\tilde{A}|\psi\rangle\rVert ^2$), then the data register is in the state $\tilde{A}|\psi\rangle /\lVert\tilde{A}|\psi\rangle\rVert$. Using amplitude estimation, one must perform this procedure an average of $1/\lVert\tilde{A}|\psi\rangle\rVert$ times for a successful measurement. Note that we use the $L^2$ norm $\lVert\cdot\rVert _2$ to quantify error as it is standard in block-encoding literature and also has relevance to linear systems applications.

Many procedures exist to generate a unitary operator which block encodes a given matrix $A$. The following theorem reduces the task of building a quantum circuit block-encoding $A$ to constructing a particular oracle operator.
\begin{theorem}
For a matrix $A$ as described above, let $O_A$ be an oracle operator defined on a $2n+1$-qubit Hilbert space:
$$O_A|0\rangle |i\rangle |j\rangle =\left( a_{ij}|0\rangle +\sqrt{1-|a_{ij}|^2}|1\rangle\right) |i\rangle |j\rangle$$
Then the operator $U_A$ described below (depicted in Figure 1) is a $\left( 1/2^{n},n+1,0\right)$-block-encoding of $A$
$$U_A=\left( I_1\otimes H^{\otimes n}\otimes I_n\right)\left( I_1\otimes\text{SWAP}\right) O_A\left( I_1\otimes H^{\otimes n}\otimes I_n\right)$$
\end{theorem}
{\bf Proof:} To prove that $U_A$ block encodes $A$, we need to prove
$$\left(\langle 0|\langle o|^{\otimes n}\langle i|\right) U_A\left( |0\rangle |0\rangle ^{\otimes n}|j\rangle\right) =\frac{1}{2^n}a_{ij}.$$
We can step through the calculations to demonstrate the validity of this equation:
\begin{align*}
\left( I_1\otimes H^{\otimes n}\otimes I_n\right)\left( |0\rangle |0\rangle ^{\otimes n}|j\rangle\right) &= \frac{1}{\sqrt{2}^n}\sum _{k=1}^n |0\rangle |k\rangle |j\rangle \\
O_A\left( I_1\otimes H^{\otimes n}\otimes I_n\right)\left( |0\rangle |0\rangle ^{\otimes n}|j\rangle\right) &= \frac{1}{\sqrt{2}^n}\sum _{k=1}^n \left( a_{kj}|0\rangle +\sqrt{1-|a_{kj}|^2}|1\rangle\right)|k\rangle |j\rangle \\
(I_1\otimes\text{SWAP})O_A\left( I_1\otimes H^{\otimes n}\otimes I_n\right)\left( |0\rangle |0\rangle ^{\otimes n}|j\rangle\right) &= \frac{1}{\sqrt{2}^n}\sum _{k=1}^n \left( a_{kj}|0\rangle +\sqrt{1-|a_{kj}|^2}|1\rangle\right)|j\rangle |k\rangle \\
\end{align*}
Meanwhile we also have
$$\left( I_1\otimes H^{\otimes n}\otimes I_n\right)\left( |0\rangle |0\rangle ^{\otimes n}|i\rangle\right) = \frac{1}{\sqrt{2}^n}\sum _{l=1}^n |0\rangle |l\rangle |i\rangle$$
and combining these two equations gives
\begin{align*}
\left(\langle 0|\langle o|^{\otimes n}\langle i|\right) U_A\left( |0\rangle |0\rangle ^{\otimes n}|j\rangle\right) &= \frac{1}{2^n}\sum _{k,l=1}^n \left(\langle 0|\langle l|\langle i|\right)\left(\left( a_{kj}|0\rangle +\sqrt{1-|a_{kj}|^2}|1\rangle\right)|j\rangle |k\rangle\right) \\
    &= \frac{1}{2^n}\sum _{k,l=1}^n a_{kj}\langle l|j\rangle\langle i|k\rangle \\
    &= \frac{1}{2^n}a_{ij}
\end{align*}
thus completing the proof. $\hfill\Box$

\begin{figure}
\[\Qcircuit @C=1em @R=0.7em {
|0\rangle && \qw & \qw & \multigate{2}{O_A} & \qw & \qw & \meter & \qw & 0 \\
|0\rangle ^{\otimes n} && \qw & \gate{H^{\otimes n}} & \ghost{O_A} & \qswap & \gate{H^{\otimes n}} & \meter & \qw & 0 \\
|\psi\rangle && \qw & \qw & \ghost{O_A} & \qswap \qwx & \qw & \qw & \frac{A|\psi\rangle}{\lVert A|\psi\rangle\rVert}
}\]
\caption{FABLE Circuit}
\end{figure}
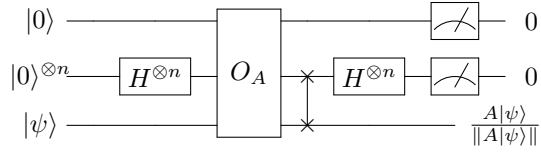

The question remains, how does one construct the oracle $O_A$? One naive implementation would be as in Figure 2, where we define the oracle to be the commuting composition of multi-qubit controlled rotations $O_A=\prod _{i,j=1}^n R_y(\theta _{ij})_{ij}$ where 
\[ R_y(\theta _{ij})_{ij}|0\rangle|i'\rangle |j'\rangle=\begin{cases} 
      \left(\cos\theta _{ij}|0\rangle +\sin\theta _{ij}|1\rangle\right) |i\rangle |j\rangle & i'=i\text{ and }j'=j \\
      |0\rangle|i'\rangle |j'\rangle & \text{otherwise}
   \end{cases}
\]
Unfortunately these multi-qubit controlled rotations are expensive to construct from one and two-qubit gates such that $O_A$ would cost $\mathcal{O}(N^4)$ of these gates to build \cite{barenco95}.

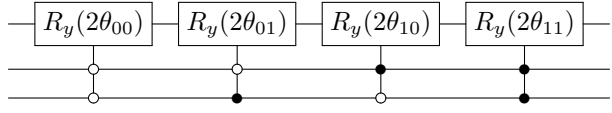
\begin{figure}
\[\Qcircuit @C=1em @R=0.7em {
& \gate{R_y(2\theta _{00})} & \gate{R_y(2\theta _{01})} & \gate{R_y(2\theta _{10})} & \gate{R_y(2\theta _{11})} & \qw \\
& \ctrlo{-1} & \ctrlo{-1} & \ctrl{-1} & \ctrl{-1} & \qw \\
& \ctrlo{-1} & \ctrl{-1} & \ctrlo{-1} & \ctrl{-1} & \qw \\
}\]
\caption{Example naive implementation for $O_A$ of a $2\times 2$ matrix $A$ using multi-qubit controlled rotations.}
\end{figure}

Instead, in a FABLE oracle we alternate single-qubit rotation gates over a \emph{different} set of angles with CNOT gates whose control qubits follow a Gray code pattern \cite{doran07} as in Figure 3. In this way, $O_A$ contains exactly $N^2$ single-qubit rotation gates and $N^2$ two-qubit CNOT gates. The following theorem describes the new set of rotation angles needed to recover the correct oracle:
\begin{theorem}
Let $\text{vec}(\theta )$ be an $N^2\times 1$ column vector with $\text{vec}(\theta )_{i+N(j-1)}=\theta _{ij}$, and let $\mbox{CNOT}^{\ i}_j$ be the CNOT gate with control qubit $i$ and target qubit $j$ (qubits numbered from 1 to $2n+1$). Then we have
$$O_A=\text{CNOT}^{\ 2}_{1}\left( R_y(\hat{\theta}_{N^2})\otimes I_{2n}\right)\left(\prod _{j=1}^{N^2-1}\text{CNOT}^{\ 2n+2-\tilde{g}_j}_{1}\left( R_y(\hat{\theta}_j)\otimes I_{2n}\right)\right)$$
where $NH^{\otimes 2n}P_G\text{vec}(\hat{\theta})=\text{vec}(\theta )$, $P_G$ permutes the $i^\text{th}$ element in $\text{vec}(\theta )$ to the $i^\text{th}$ Gray code element $g_i$, the product denotes left-multiplication and $\tilde{g}_j$ is the digit corresponding to the bit flip between $g_{j}$ and $g_{j-1}$, the $(j-1)^\text{th}$ and $j^\text{th}$ Gray code numbers.  
\end{theorem}
{\bf Proof:} Notice that due to the structure of this circuit, for every fixed basis state of the last $2n$ qubits, the action on the first qubit is given by a product of rotation gates $R_y(\hat{\theta}_j)$ with $X$ gates interspersed in some pattern depending on the basis state. Recall the identities $R_y(\hat{\theta}_i)R_y(\hat{\theta}_j)=R_y(\hat{\theta}_i+\hat{\theta}_j)$ and $XR_y(\hat{\theta}_i)X=R_y(-\hat{\theta}_i)$. Thus, for a basis state $i$, the rotation angle $\hat{\theta}_j$ flips parity when the binary vector $b_i$ has a 1 in the $\tilde{g_j}$-digit. This implies the existence of a matrix $M$ with entries $\pm 1$ which satisfies $M\text{vec}(\hat{\theta} )=\text{vec}(\theta )$ and using these identities we may write $(M)_{ij}=(-1)^{b_i\cdot g_j}$ where the dot product is over binary vectors. Recognizing that $(H^{\otimes n})_{ij}=\frac{1}{\sqrt{N}}(-1)^{b_i\cdot b_j}$ gives the result. $\hfill\Box$

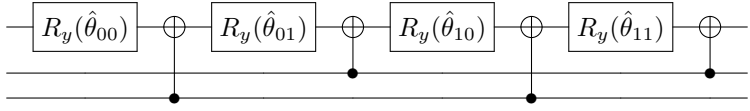
\begin{figure}
\[\Qcircuit @C=1em @R=0.7em {
& \gate{R_y(\hat{\theta} _{00})} & \targ & \gate{R_y(\hat{\theta} _{01})} & \targ & \gate{R_y(\hat{\theta} _{10})} & \targ & \gate{R_y(\hat{\theta} _{11})} & \targ & \qw \\
& \qw & \qw & \qw & \ctrl{-1} & \qw & \qw & \qw & \ctrl{-1} & \qw \\
& \qw & \ctrl{-2} & \qw & \qw & \qw & \ctrl{-2} & \qw & \qw & \qw \\
}\]
\caption{FABLE implementation for $O_A$ of a $2\times 2$ matrix $A$ alternating single-qubit rotations with CNOT gates. Extending this circuit to block-encoding $N\times N$ matrices would take exactly $N^2$ rotations and $N^2$ CNOT gates. Notice that we use the transformed angles $\hat{\theta}$ in these rotation gates.}
\end{figure}

These two theorems completely describe the FABLE circuit. In particular, a full FABLE circuit contains $N^2$ single-qubit rotation gates, $N^2+3n$ two-qubit CNOT gates, and $2n$ Hadamard gates. For block encoding a dense matrix with $N^2$ unique elements this is close to ideal. However, one may ask whether this circuit can be shortened with a minimal impact on accuracy. Camps and Van Beeuman in \cite{fable22} introduce the `approximate' component of FABLE by removing rotation gates if the angles lie below a certain threshold, otherwise if $|\hat{\theta}_j|<\delta$. This not only reduces rotation gate count, but potentially CNOT count as well since CNOT gates are both involutory and can commute with CNOTs of the same target, as shown in Figure 4. Let $O_A(\delta)$ be the oracle given by eliminating rotation gates below a threshold $\delta$. An extremely loose upper bound on the accuracy of these approximate FABLE circuits is given in \cite{fable22}:
\begin{proposition}
The operator $U_A(\delta )$ below is a $(1/2^n,n+1,N^3\delta )$ block encoding of $A$:
$$U_A(\delta )=\left( I_1\otimes H^{\otimes n}\otimes I_n\right)\left( I_1\otimes\text{SWAP}\right) O_A(\delta )\left( I_1\otimes H^{\otimes n}\otimes I_n\right)$$
\end{proposition}
Based on figures provided in their paper, the error on this block encoding in practice seems to be orders of magnitude less than the bound provided by proposition 1.

\begin{figure}
Complete oracle circuit:
\[\Qcircuit @C=1em @R=0.7em {
& \gate{\hat{\theta} _{0}} & \targ & \gate{\hat{\theta} _{1}} & \targ & \gate{\hat{\theta} _{2}} & \targ & \gate{\hat{\theta} _{3}} & \targ & \gate{\hat{\theta} _{4}} & \targ & \gate{\hat{\theta} _{5}} & \targ & \gate{\hat{\theta} _{6}} & \targ & \gate{\hat{\theta} _{7}} & \targ & \qw \\
& \qw & \qw & \qw & \qw & \qw & \qw & \qw & \ctrl{-1} & \qw & \qw & \qw & \qw & \qw & \qw & \qw & \ctrl{-1} & \qw \\
& \qw & \qw & \qw & \ctrl{-2} & \qw & \qw & \qw & \qw & \qw & \qw & \qw & \ctrl{-2} & \qw & \qw & \qw & \qw & \qw \\
& \qw & \ctrl{-3} & \qw & \qw & \qw & \ctrl{-3} & \qw & \qw & \qw & \ctrl{-3} & \qw & \qw & \qw & \ctrl{-3} & \qw & \qw & \qw 
}\]

Removing rotations below threshold:
\[\Qcircuit @C=1em @R=0.7em {
& \gate{\hat{\theta} _{0}} & \targ & \gate{\hat{\theta} _{1}} & \targ & \targ & \targ & \targ & \targ & \targ & \gate{\hat{\theta} _{7}} & \targ & \qw \\
& \qw & \qw & \qw & \qw & \qw & \ctrl{-1} & \qw & \qw & \qw & \qw & \ctrl{-1} & \qw \\
& \qw & \qw & \qw & \ctrl{-2} & \qw & \qw & \qw & \ctrl{-2} & \qw & \qw & \qw & \qw \\
& \qw & \ctrl{-3} & \qw & \qw & \ctrl{-3} & \qw & \ctrl{-3} & \qw & \ctrl{-3} & \qw & \qw & \qw 
}\]

Commuting and eliminating CNOT gates:
\[\Qcircuit @C=1em @R=0.7em {
& \gate{\hat{\theta} _{0}} & \targ & \gate{\hat{\theta} _{1}} & \targ & \targ & \gate{\hat{\theta} _{7}} & \targ & \qw \\
& \qw & \qw & \qw & \qw & \ctrl{-1} & \qw & \ctrl{-1} & \qw \\
& \qw & \qw & \qw & \qw & \qw & \qw & \qw & \qw \\
& \qw & \ctrl{-3} & \qw & \ctrl{-3} & \qw & \qw & \qw & \qw 
}\]
\caption{Illustration of the compression of an 8-element uniform controlled rotation. If angles $\{\hat{\theta}_k\} _{k\in\{ 2,...,6\}}$ lie below a threshold $\delta$, then they can be eliminated from the circuit (middle) and the intermediate CNOTs can commute and cancel (bottom). This particular compression eliminates 5 rotation gates and 4 CNOT gates.}
\end{figure}

\section{S-FABLE: Modifying FABLE for unstructured sparse matrices}\label{SFABLE}

In \cite{fable22}, Camps and Van Beeuman posed the question, what kinds of matrices and problem domains lead to high compression in FABLE circuits? While it was mentioned that matrices $A$ sparse under a Hadamard-Walsh transform empirically seem to compress well, the mapping from a matrix $A$ into a set of angles $\text{vec}(\hat{\theta})$ is nonlinear and its inverse is difficult to globally characterize.

Rather than lead with a solution in search of a problem, we instead ask the converse; is there a way to modify the standard FABLE algorithm to accommodate application-relevant matrices? One obvious place to start would be with sparse matrices as they are well-represented in a myriad of quantum algorithms requiring block-encoding \cite{fang23}. Specifically, if the number of nonzero entries of a matrix scales at $\mathcal{O}(N)$ rather than $\mathcal{O}(N^2)$ for generic dense matrices, we ideally would seek a block-encoding circuit which also contains $\mathcal{O}(N)$ one and two-qubit gates. Since sparse matrices are in general dense in the Hadamard-Walsh domain (i.e. $H^{\otimes n}AH^{\otimes n}$ is dense), their FABLE circuits do not have advantageous scaling. For the remainder of the paper we will abbreviate $H^{\otimes n}$ as $H$ for readability when not ambiguous.

Though a sparse matrix $A$ is generally dense in the Hadamard-Walsh domain, because the Hadamard-Walsh transform is involutory the matrix $HAH$ is sparse in the transformed domain (i.e. $H(HAH)H=A$ is sparse). Thus we may expect the oracle $O_{HAH}$ to compress well, although $U_{HAH}$ now block encodes the transformed matrix $HAH$. We must be careful however to ensure that $\lVert HAH\rVert _\infty \le 1$, otherwise the oracle is not well defined; we instead consider the oracle $O_{HAH/\lVert HAH\rVert _\infty}$. To recover the original matrix, we use the following observation:
\begin{lemma}
Let $U_A$ be a $(\alpha ,a,\epsilon )$ block-encoding of $A$, and let $U_1,U_2$ be $n$-qubit matrices of the same size as $A$. Then $U=(U_2\otimes I_{2^a})U_A(U_1\otimes I_{2^a})$ is a $(\alpha ,a,\epsilon )$ block-encoding of $U_2AU_1$.
\end{lemma}
{\bf Proof:} That $U$ block-encodes $U_2AU_1$ is immediately apparent from the matrix representation:
$$U=\begin{pmatrix} U_2 & ~ \\ ~ & \ddots\end{pmatrix}\begin{pmatrix} \alpha A & \cdot \\ \cdot & \cdot\end{pmatrix}\begin{pmatrix} U_1 & ~ \\ ~ & \ddots\end{pmatrix}$$
That the error bound $\epsilon$ still holds follows from the fact that unitary operators are isometries on $L^2$. $\hfill\Box$

This lemma along with the procedure described allows us to formally describe the Sparse-FABLE (S-FABLE) algorithm.
\begin{theorem}
Let $U_A$ be a $(1/2^n,n+1,0)$ block encoding of $A$ as defined in Proposition 1. Then the operator
$$U^s_A=(H^{\otimes n}\otimes I_{2^{n+1}})U_{HAH/\lVert HAH\rVert _\infty}(H^{\otimes n}\otimes I_{2^{n+1}})$$
(depicted in Figure 5) is a $(1/(2^n\lVert HAH\rVert _\infty ),n+1,0)$ block encoding of $A$.
\end{theorem}

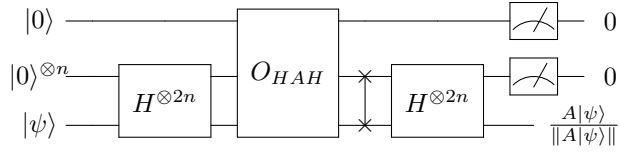
\begin{figure}
\[\Qcircuit @C=1em @R=0.7em {
|0\rangle && \qw & \qw & \multigate{2}{O_{HAH}} & \qw & \qw & \meter & \qw & 0 \\
|0\rangle ^{\otimes n} && \qw & \multigate{1}{H^{\otimes 2n}} & \ghost{O_{HAH}} & \qswap & \multigate{1}{H^{\otimes 2n}} & \meter & \qw & 0 \\
|\psi\rangle && \qw  & \ghost{H^{\otimes 2n}} & \ghost{O_{HAH}} & \qswap \qwx &  \ghost{H^{\otimes 2n}} & \qw & \frac{A|\psi\rangle}{\lVert A|\psi\rangle\rVert}
}\]
\caption{S-FABLE Circuit}
\end{figure}

The mechanics of the S-FABLE algorithm are not so complex, one simply block-encodes a transformed operator $HAH$ with FABLE and undoes the transform by surrounding the FABLE circuit by Hadamard gates. While for general matrices $A$ the S-FABLE circuit is of no immediate use, if $A$ is sparse one is able to leverage a highly compressible FABLE oracle $O_{HAH}$ at the cost of $2n$ extra Hadamard gates as well as an additional $\mathcal{O}(N^2\log{N})$ classical operations to compute $HAH$ \cite{beer81}.

\section{LS-FABLE: Eliminating classical overhead on S-FABLE}\label{LSFABLE}

While S-FABLE provides an intriguing alternative to the FABLE method for sparse matrices, it still retains the $\mathcal{O}(N^2\log{N})$ classical overhead required to conduct several matrix multiplications by $H^{\otimes n}$ as well as computing inverse cosine for each element of the generally dense matrix $HAH$. Again, as was argued with the quantum resources, ideally we would utilize only $\mathcal{O}(N)$ classical resources for data that contains $\mathcal{O}(N)$ nonzero elements.

This motivates a modification of S-FABLE we call `Lazy' S-FABLE (LS-FABLE), that eliminates this quadratic overhead and instead uses exactly the number of classical operations that there are nonzero entries in the matrix to be block-encoded $A$. The LS-FABLE circuit has a similar construction to the S-FABLE circuit. However, we make the ambitious assumption that for the sparse matrices we are working with enough entries of $\frac{1}{N}HAH$ satisfy $\cos ^{-1}(x)\approx\frac{\pi}{2}-x$. Under this assumption the matrix of transformed angle matrix becomes $\hat{\theta}=\frac{\pi}{2}E_{1,1}-\frac{1}{N}A$ where $E_{1,1}$ is the matrix with a single nonzero element 1 in the top-left corner. If we use this matrix of transformed angles in our oracle, then the resulting circuit precisely block encodes $H\sin (HAH)H$ (sine taken to be an entrywise operation) which we claim is approximately $A$ for a broad class of sparse matrices.

The LS-FABLE method has several advantages over FABLE and S-FABLE, notably that the classical resources needed to calculate the angles used in the rotation gates amount to simply dividing each nonzero entry of $A$ by $N$. Further, the number of rotation gates in the oracle is fixed at the number of nonzero elements in $A$; both of these quantum and classical costs are small (and easily quantifiable) if $A$ is sparse. However, making use of these brazen approximations can come at the cost of accuracy, and in LS-FABLE that accuracy is fixed for each matrix $A$ since higher order approximations of $\cos^{-1}(HAH)$ are not sparse in general. Fortunately, in the next section we will find many cases where the accuracy of LS-FABLE is commensurate with that of S-FABLE.

\section{Numerical Results}\label{DATA}

In this section we compare the performance of the three algorithms presented in a myriad of instances. We find that S-FABLE and LS-FABLE outperform FABLE for sparse matrices with little structure. As the matrices exhibit more structure, FABLE improves comparatively.

Because LS-FABLE is a static method not parameterizable by accuracy, we take care to explicitly define our axes of comparison. If $A$ is the matrix to be block-encoded, let $n_F(A,\epsilon )$ be the total number of rotation gates needed for the FABLE approximation of $A$ to satisfy $\lVert A-A_F\rVert _2<\epsilon$. Define a similar function $n_S(A,\epsilon )$ for S-FABLE. When comparing FABLE and S-FABLE, these quantities will be our primary focus.

If we aim to compare all three of FABLE, S-FABLE, and LS-FABLE, let $|A|$ be the number of nonzero entries in $A$ such that the LS-FABLE block encoding of $A$ uses exactly $|A|$ rotation gates. Then let $\epsilon _F(A)$ be the error in the FABLE approximation of $A$ using only $|A|$ rotation gates, and define similar quantities $\epsilon _S(A)$ and $\epsilon _{LS}(A)$. Since LS-FABLE uses a fixed number of rotation gates for matrices of specified sparsity, we compare the $L^2$ error across methods for the fixed rotation gate count. We could also compare gate counts used by the different methods to achieve the fixed LS-FABLE accuracy $\epsilon _{LS}(A)$.

Rather than use the traditional measure of sparsity $s=|A|/N^2$ as the fraction of nonzero entries in a given matrix, we will instead focus on matrices with a fixed \emph{relative sparsity} $s=|A|/N$, or average number of nonzero entries per row/column. These matrices have $\mathcal{O}(N)$ nonzero entries (e.g. representing graphs of variable size but fixed connectivity) and allow us to make better observations about asymptotic scaling.

\subsection{Uniform Random Sparse Data}

We first begin by analyzing performance of these three algorithms on the most favorable instances for the new sparse methods, uniform random sparse data, or matrices with a number of nonzero entries scaling as $\mathcal{O}(N)$, and each nonzero entry uniformly distributed from $[-1,1]$. In particular, we empirically demonstrate that both S-FABLE and LS-FABLE exhibit exponential improvement over FABLE in either accuracy or circuit size as the number of qubits increases. For example, Table 1 lists one and two-qubit gate counts for FABLE and S-FABLE on a 13 qubit block with 12 nonzero entries per row encoded to an $L^2$ norm accuracy of $\epsilon =2^{-10}$; while FABLE required a nearly complete circuit, S-FABLE used just under 1.6\% of the number of total gates used by FABLE.

\begin{table}
\begin{center}
\begin{tabular}{ |p{2cm}||p{2cm}|p{2cm}|  }
 \hline
 ~ & FABLE & S-FABLE \\
 \hline
 \# Rotation & 66,823,419 & 98,232 \\
 \hline
 \# CNOT & 67,107,709 & 543,713 \\
 \hline
 \# Had & 26 & 52 \\
 \hline
 {\bf Total} & {\bf 133,931,154} & {\bf 641,997} \\
 \hline
\end{tabular}
\caption{Performance comparison of FABLE vs S-FABLE for a random matrix with $n=13$, $s=12$, and $\epsilon =2^{-10}$.}
\end{center}
\end{table}

For a fixed error $\epsilon$ and a fixed sparsity $s$, as the number of qubits increases S-FABLE appears to have exponentially increasing benefit over FABLE in the number of gates required to perform the approximation to $\epsilon$ accuracy (more on this later). Figure 6 depicts FABLE circuits for generic sparse matrices as affording hardly any compression, while past a threshold number of qubits, S-FABLE begins to perform orders of magnitude better than FABLE. This threshold number of qubits before S-FABLE delivers advantage is dependent on the relative sparsity $s$; denser matrices require a higher qubit count before S-FABLE requires less gates than FABLE.

\begin{figure}
\centering
\includegraphics[scale=0.5]{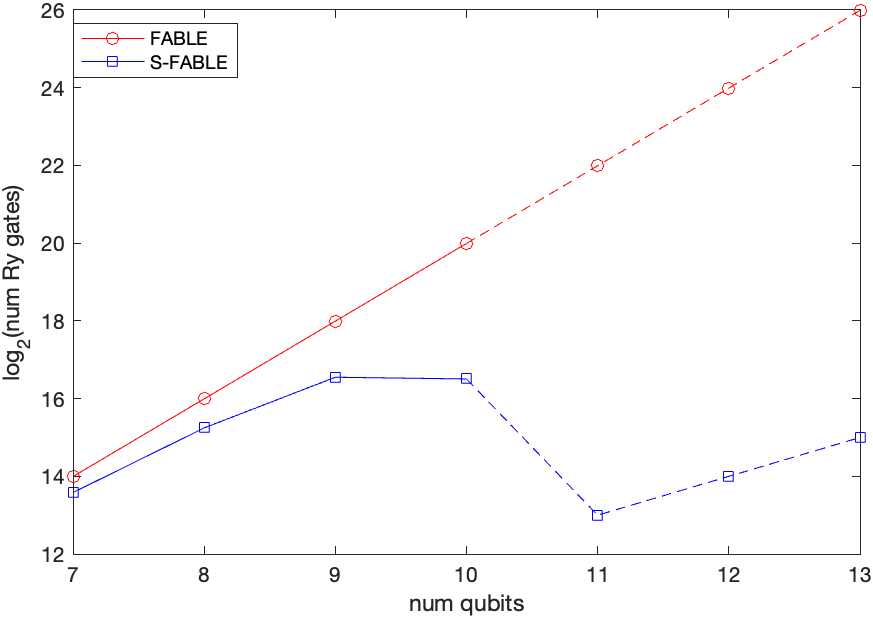}\hspace{0.8cm}\includegraphics[scale=0.5]{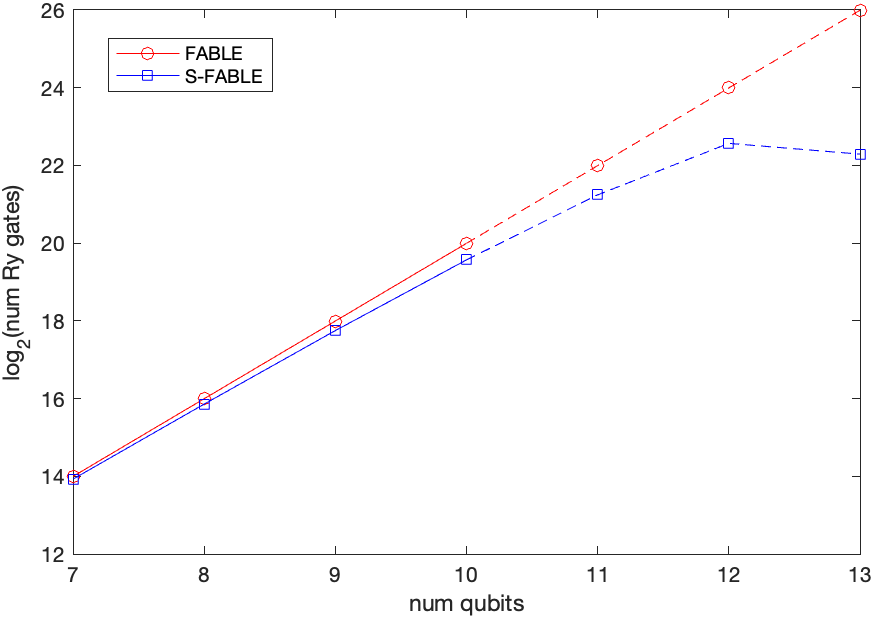}
\caption{Comparison of $n_F(A,\epsilon )$ and $n_S(A,\epsilon )$ with $\epsilon =2^{-10}$ and (\emph{left}) $s=4$ and (\emph{right}) $s=16$. Quantities for $n\in\{ 7,...,10\}$ averaged over 100 samples of $A$, while only 1 sample was used for $n\in\{ 11,12,13\}$.}
\end{figure}

While $n_F(A,\epsilon )$ appears to be close to the maximum FABLE circuit size for general sparse matrices $A$ (the FABLE circuit in Table 1 uses 99.57\% of the maximum $2^{2n}$ rotation gates and 99.998\% of the maximum number of CNOT gates in the oracle), Figure 7 illustrates a different story about the dependence of $n_S(A,\epsilon )$ on the $L^2$ error bound $\epsilon$. In particular, for a given sparse matrix $A$, there appears to be a threshold value $\tilde{\epsilon}$ which, for $\epsilon >\tilde{\epsilon}$, $n_S(A,\epsilon )$ appears to be nearly constant and orders of magnitude smaller than the maximal gate count of $n_F(A,\epsilon )$. Conversely, as $\epsilon\rightarrow 0$ past the threshold value $\tilde{\epsilon}$, $n_S(A,\epsilon )$ approaches its maximal gate count at a value of $n_F(A,\epsilon )$. As $n$ increases, it appears that this threshold value $\tilde{\epsilon}$ decreases exponentially (supported by later arguments).

\begin{figure}
\centering
\includegraphics[scale=0.7]{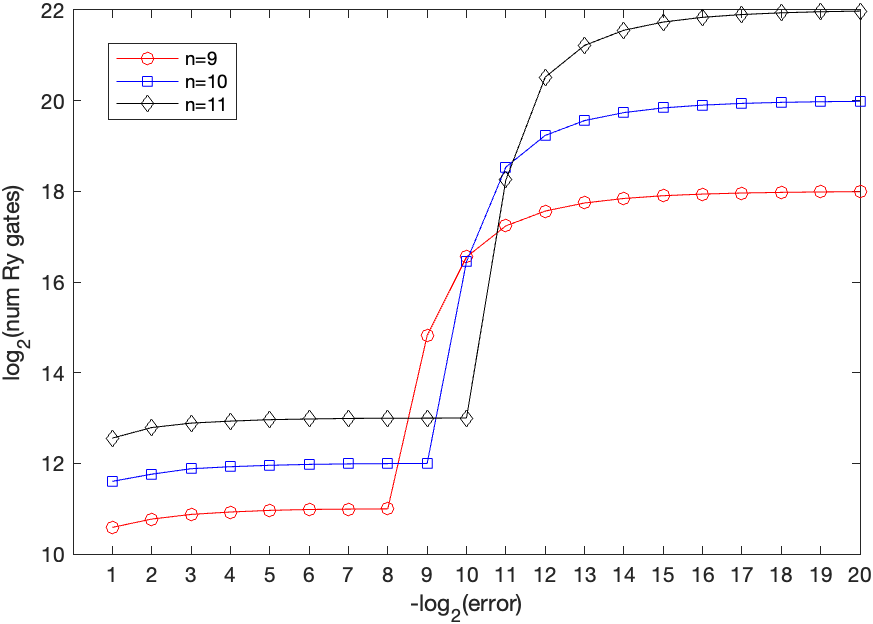}
\caption{Comparison of $n_S(A_k,\epsilon )$ where $A_k$ is a $2^k\times 2^k$ sparse matrix with $k\in\{ 9,10,11\}$ (blue, red, and black lines respectively), $\epsilon\in\{ 2^{-1},...,2^{-20}\}$, and $s=4$.}
\end{figure}

The quantity $n_S(A,\epsilon )$ behaves similarly with variable sparsity and all other parameters fixed. Namely, there appears to be a threshold sparsity $\tilde{s}$ such that $n_S(A,\epsilon )$ is nearly constant for all $s<\tilde{s}$ and approaches the upper maximum limit for $s>\tilde{s}$. As the number of qubits increases, this threshold quantity $\tilde{s}$ also increases.

\begin{figure}
\centering
\includegraphics[scale=0.7]{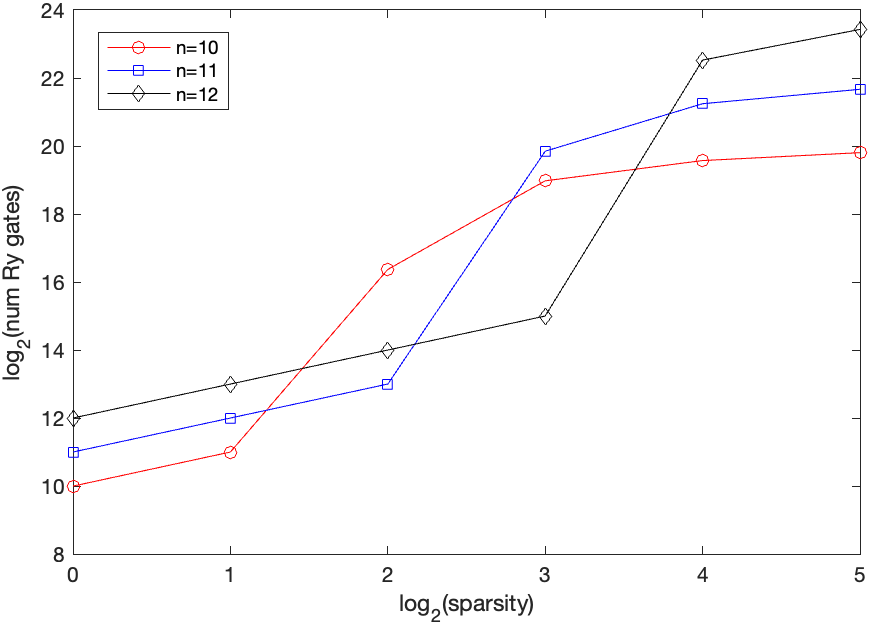}
\caption{Comparison of $n_S(A_k,\epsilon )$ where $A_k$ is a $2^k\times 2^k$ sparse matrix with $k\in\{ 10,11,12\}$ (blue, red, and black lines respectively), $\epsilon =2^{-10}$, and $s\in\{ 2^0,...,2^5\}$.}
\end{figure}

LS-FABLE exhibits similar improvements as S-FABLE over FABLE. Figure 9 depicts a plot of errors achieved by the block-encodings when restricted to the fixed $|A|$ number of rotation gates used in LS-FABLE. While FABLE performs poorly when its rotation gates are restricted in this way ($\epsilon _F(A)\approx \mathcal{O}(1)$), both S-FABLE and LS-FABLE appear to become exponentially more accurate over fixed sparsity as the number of qubits increases. From Figure 10 both $\epsilon _S(A)$ and $\epsilon _{LS}(A)$ appear to increase exponentially as sparsity increases. Using a linear regression over $n$ and $s$ (as they both appear linearly related to $\epsilon$ on a log scale), we estimate the errors achieved by S-FABLE and LS-FABLE to be as follows:
\begin{equation}\label{eq}
    \epsilon _{S}(A)\approx 0.3087\cdot\frac{s^{1.4634}}{N^{1.0778}},\hspace{1cm}\epsilon _{LS}(A)\approx 0.2969\cdot\frac{s^{1.6709}}{N^{1.0191}}
\end{equation}
Notably, $\epsilon _{S}(A)$ offers a nearly $\mathcal{O}(1)$ factor of improvement in $N$ over $\epsilon _{LS}(A)$ (i.e. $\epsilon _S/\epsilon _{LS}\approx \mathcal{O}(1/N^{0.06})$, although S-FABLE is comparatively more robust to increases in matrix density than LS-FABLE is. Furthermore, these approximations demonstrate that the sparse versions of FABLE still approximately improve in accuracy as the system size increases as long as $s<\mathcal{O}(N^{0.6})$.

\begin{figure}
\centering
\includegraphics[scale=0.5]{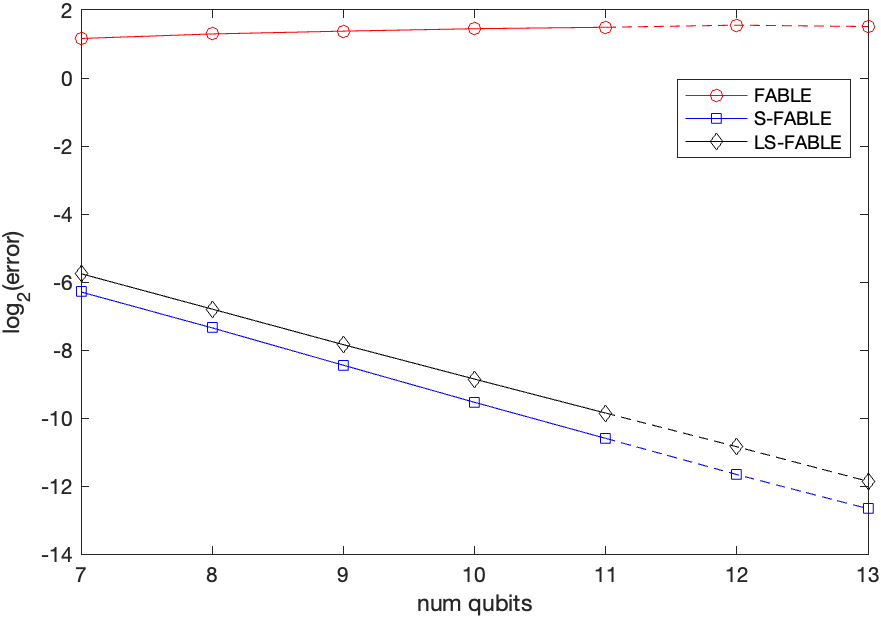}\hspace{0.8cm}\includegraphics[scale=0.5]{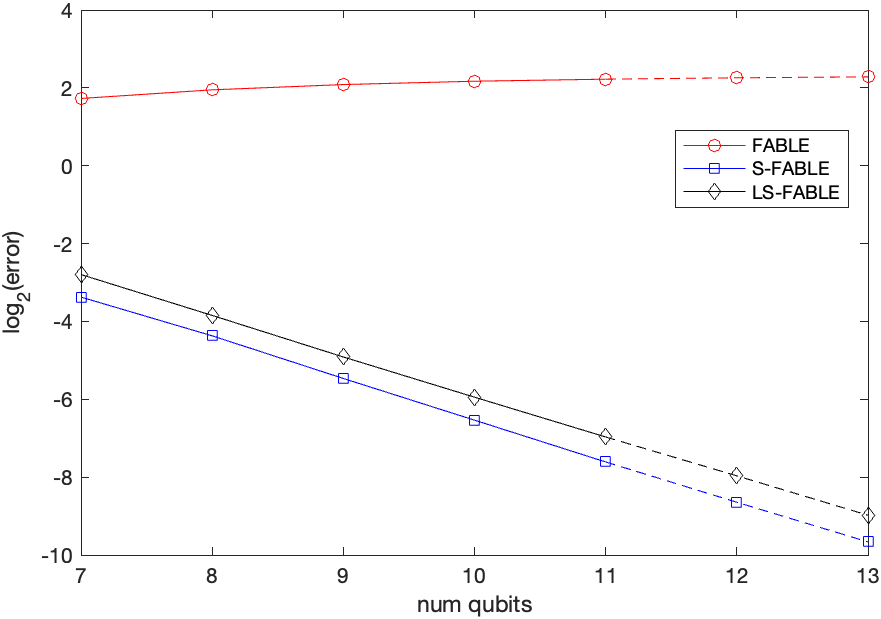}
\caption{Comparison of $\epsilon _F(A)$, $\epsilon _S(A)$, and $\epsilon _{LS}(A)$ over $n\in\{ 7,...,13\}$ with (\emph{left}) $s=4$ and (\emph{right}) $s=16$. Quantities for $n\in\{ 7,...,11\}$ averaged over 100 samples of $A$, while only 1 sample was used for $n=12$ and $n=13$.}
\end{figure}

\begin{figure}
\centering
\includegraphics[scale=0.5]{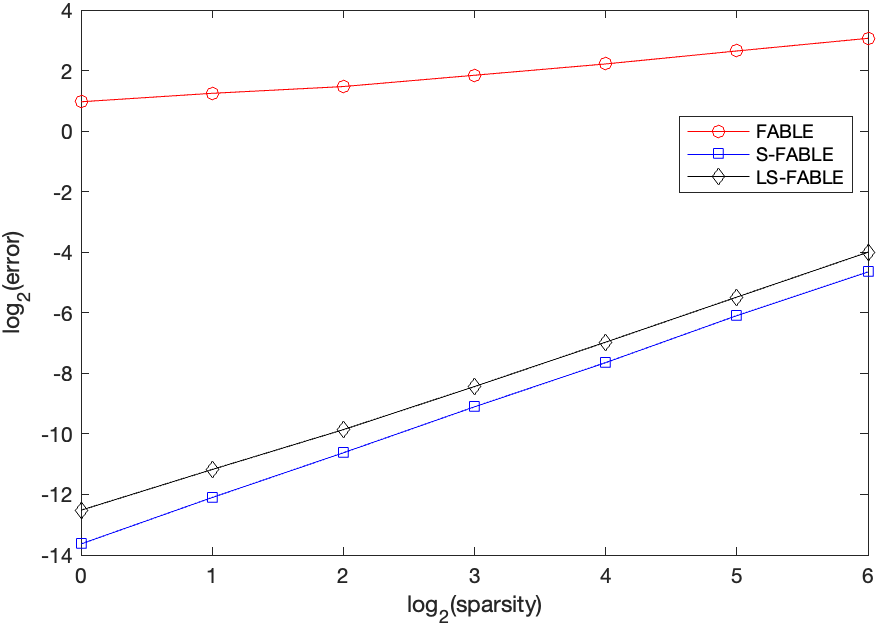}\hspace{0.8cm}\includegraphics[scale=0.5]{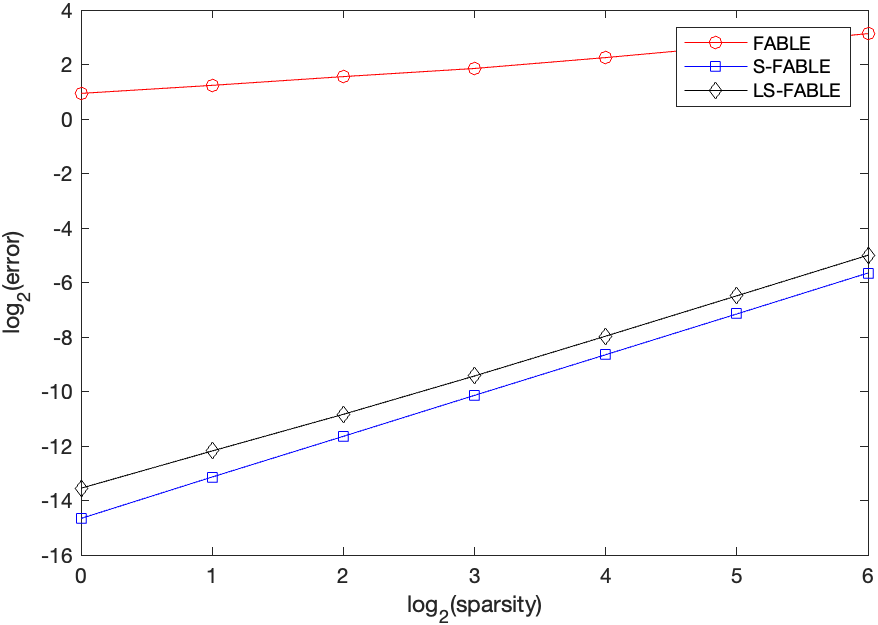}
\caption{Comparison of $\epsilon _F(A)$, $\epsilon _S(A)$, and $\epsilon _{LS}(A)$ over $s\in\{ 2^0,...,2^6\}$ with (\emph{left}) $n=11$ and (\emph{right}) $n=12$.}
\end{figure}

\subsection{Physical Systems}

While from the results contained previous section, S-FABLE and LS-FABLE appear to exponentially outperform FABLE as the size of the unstructured sparse matrix increases, the same cannot be said if the sparse matrix contains symmetries or structure. We revisit a few of the examples of block-encodings given in \cite{fable22}.

First, we analyze the following Heisenberg spin chain Hamiltonian \cite{joyce67} given by
$$H=\sum _{i=1}^{n-1}\left(J_xX^{(i)}X^{(i+1)}+J_yY^{(i)}Y^{(i+1)}+J_zZ^{(i)}Z^{(i+1)}\right) +\sum _{i=1}^n h_zZ^{(i)}$$
where $X^{(i)}$ is the Pauli $X$ gate on the $i^\text{th}$ qubit and likewise for $Y^{(i)}$ and $Z^{(i)}$. While these Hamiltonians are sparse (i.e. $s=\log{N}$) the S-FABLE and LS-FABLE circuits entirely fail at delivering highly compressed block-encoding circuits the way they succeeded in the case of unstructured matrices. Figure 11 illustrates that FABLE performs nearly identically to S-FABLE in the case of a Heisenberg XXX model (i.e. $J_x=J_y=J_z$ and $h_z=0$), the most symmetric of Heisenberg Hamiltonians. Meanwhile, from observing Figure 11 and Table 2, S-FABLE and LS-FABLE perform comparatively better for a Heisenberg XYZ model with parameters taking on random values, yet $\epsilon _S(H)$ and $\epsilon _{LS}(H)$ appear to limit to a constant value and do not achieve the exponential scaling observed in Figure 9; equation \hyperref[eq]{(1)} would predict that if $A$ was a 13-qubit matrix with sparsity $s=13$, we would have $\epsilon _{LS}(A)\approx 2^{-8.57}$ rather than the value of $\approx 2^{-2.7}$ we see here.

\begin{table}
\begin{center}
\begin{tabular}{ |p{2cm}||p{2cm}|p{2cm}|  }
 \hline
 ~ & FABLE & S-FABLE \\
 \hline
 \# Rotation & 16,685,043 & 16,685,038 \\
 \hline
 \# CNOT & 50,344,375 & 33,484,797 \\
 \hline
 \# Had & 26 & 52 \\
 \hline
 {\bf Total} & {\bf 67,029,444} & {\bf 50,169,887} \\
 \hline
\end{tabular}
\caption{Performance comparison of FABLE vs S-FABLE for a Heisenberg XXX model ($J_x=J_y=J_z$, $h_z=0$) with $n=13$ and $\epsilon =2^{-10}$.}
\end{center}
\end{table}

\begin{figure}
\centering
\includegraphics[scale=0.5]{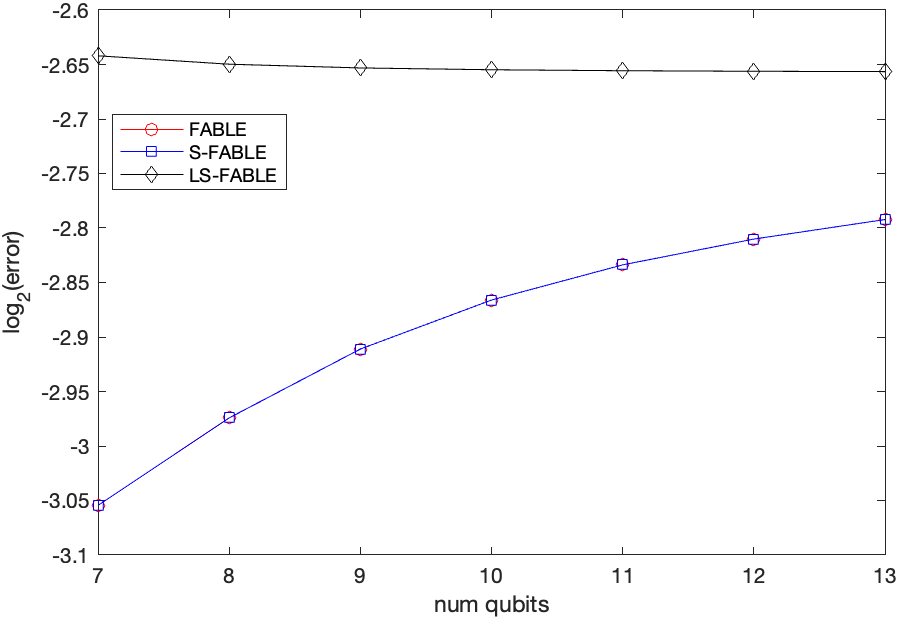}\hspace{0.8cm}\includegraphics[scale=0.5]{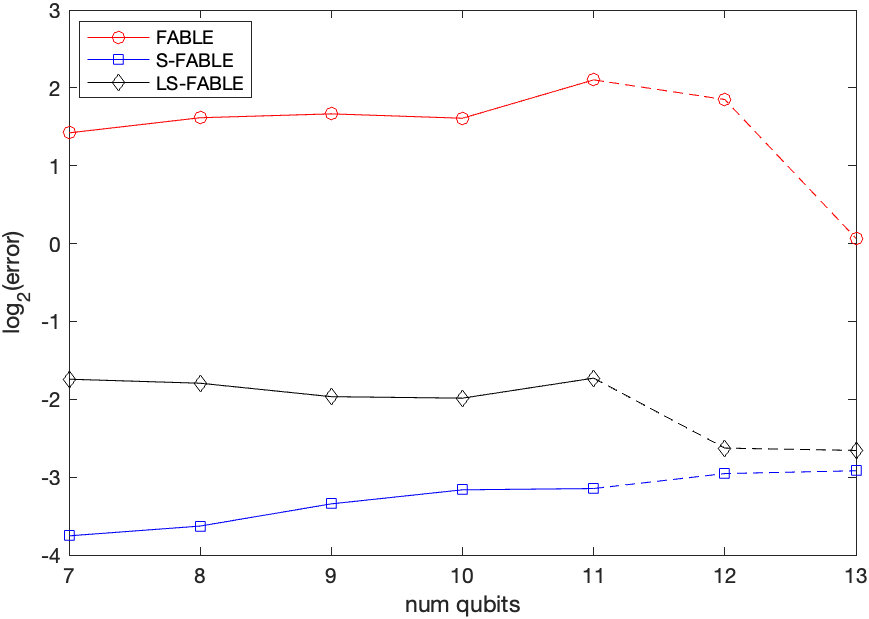}
\caption{Comparison of $\epsilon _F(H)$, $\epsilon _S(H)$, and $\epsilon _{LS}(H)$ with $H$ as a Heisenberg Hamiltonian (\emph{left}) $H$ is a Heisenberg $XXX$ Hamiltonian with $J_x=J_y=J_z$ and $h_z=0$ (\emph{right}) $H$ is a Heisenberg $XYZ$ Hamiltonian with quantities $J_x,J_y,J_z,h_z$ uniformly sampled from the interval $[-1,1]$. Error quantities are averaged over 100 samples for each $n\in\{ 6,...,11\}$ while a single sample is used for each $n=12,13$.}
\end{figure}

The situation is similar for Laplacians arising in differential equations \cite{evans10}. Suppose $L_{xx}$ is a discretized 1D Laplace operator matrix with $(L_{xx})_{ii}=2$ and $(L_{xx})_{ij}=-1$ when $|i-j|=1$. Using this, we can define a 2D Laplace operator of the form $L=L_{xx}\otimes I+I\otimes L_{yy}$ where $L_{xx}$ and $L_{yy}$ are of different dimensions. While these Laplacian matrices are very sparse (i.e. $s=5$), their symmetry causes S-FABLE and LS-FABLE to perform poorly. In Figure 12, we see that the sparse methods do not perform or scale to the expectation of \hyperref[eq]{(1)}.

\begin{figure}
\centering
\includegraphics[scale=0.5]{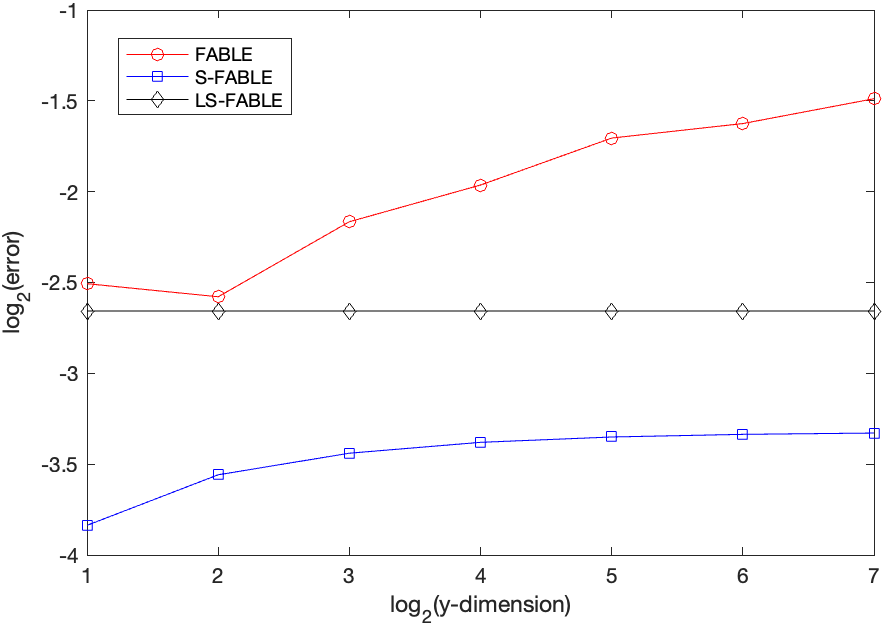}\hspace{0.8cm}\includegraphics[scale=0.5]{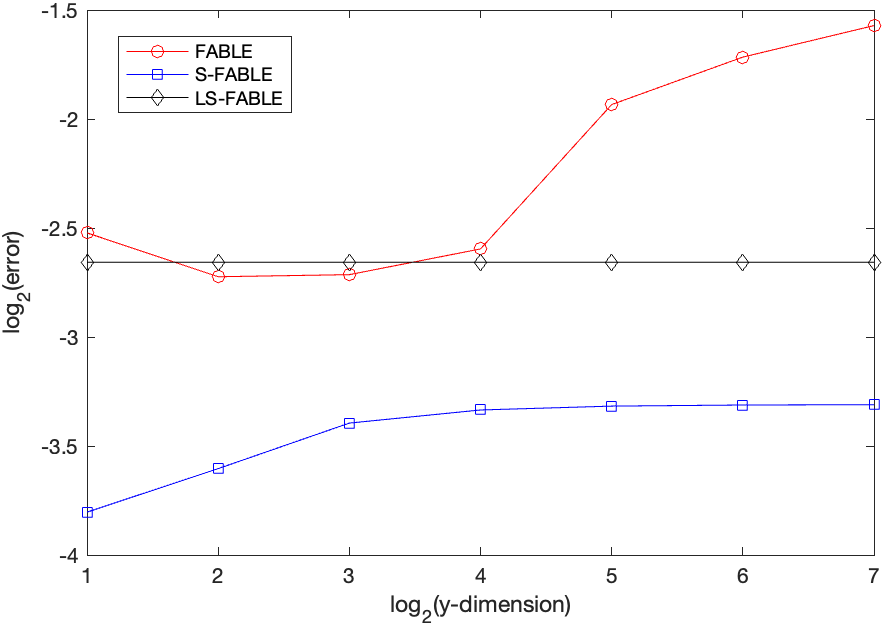}
\caption{Comparison of $\epsilon _F(L)$, $\epsilon _S(L)$, and $\epsilon _{LS}(L)$ with $L$ as a 2D Laplace operator with $x$-dimension $2^6$ (\emph{left}) non-periodic boundary condition (\emph{right}) periodic boundary condition}
\end{figure}

While for these highly symmetric and structured sparse matrices we find FABLE achieves relatively higher compression and accuracy compared to the sparse methods, alternative bespoke methods exist to block-encode the matrices in question \cite{low19}. For example, a Heisenberg XYZ model only takes 4 parameters, it seems highly inefficient to use LS-FABLE to encode $N\log{N}$ rotation gates in its oracle as opposed to using LCU-based methods \cite{childs12}.

\subsection{Positive Data}

We just demonstrated that S-FABLE and LS-FABLE forfeit their advantage over FABLE when the matrices being block-encoded have some symmetry as in the case of certain physical systems like Heisenberg Hamiltonians and PDE Laplacians. However this is not the only subset of matrices that the sparse methods fail to efficiently block-encode. In section, we demonstrate that S-FABLE and LS-FABLE also perform poorly when the matrices are nonnegative. Figure 13 and Table 3 illustrate this to be the case; when data is either binary (e.x. graph adjacency matrices) or otherwise nonnegative, both $\epsilon _S(A)$ and $\epsilon _{LS}(A)$ appear to limit to an $O(1)$ quantity rather than exhibit exponential scaling for sparse matrices with both positive and negative nonzero values. Depending on the application, this shortcoming may be overcome by a prior classical transformation on the data.

\begin{table}
\begin{center}
\begin{tabular}{ |p{2cm}||p{2cm}|p{2cm}|  }
 \hline
 ~ & FABLE & S-FABLE \\
 \hline
 \# Rotation & 65,929,353 & 66,711,239 \\
 \hline
 \# CNOT & 67,087,599 & 67,106,583 \\
 \hline
 \# Had & 26 & 52 \\
 \hline
 {\bf Total} & {\bf 133,016,978} & {\bf 133,817,874} \\
 \hline
\end{tabular}
\caption{Performance comparison of FABLE vs S-FABLE for a random matrix with all nonzero entries equal to 1, with $n=13$, $s=12$, and $\epsilon =2^{-10}$.}
\end{center}
\end{table}

\begin{figure}
\centering
\includegraphics[scale=0.5]{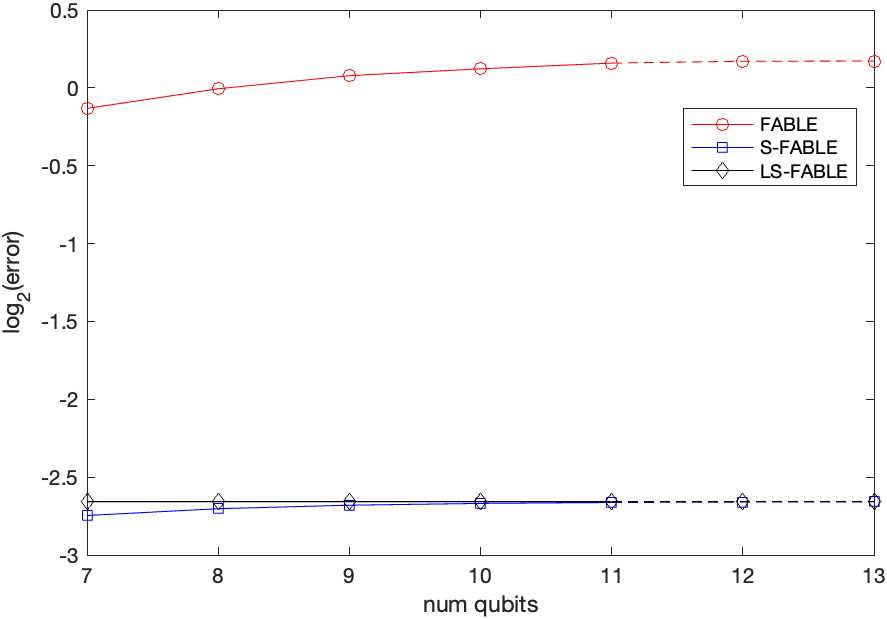}\hspace{0.8cm}\includegraphics[scale=0.5]{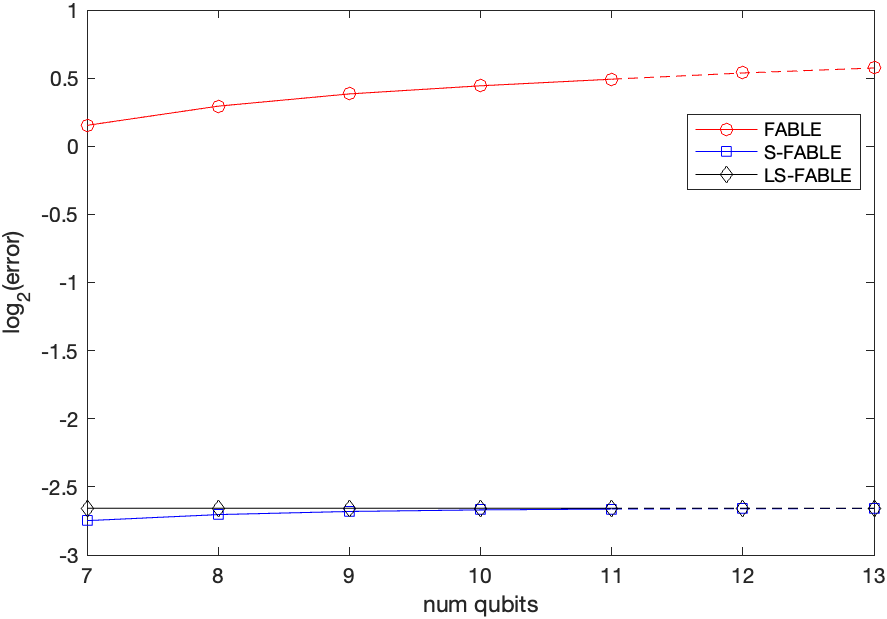}
\caption{Comparison of $\epsilon _F(A)$, $\epsilon _S(A)$, and $\epsilon _{LS}(A)$ with $A$ as an unstructured random matrix with $s=4$ and $n\in\{ 7,...,13\}$. For $n\in\{ 7,...,11\}$, error quantities are averaged over 100 samples of $A$, while for each $n=12,13$ only one sample of $A$ is used (\emph{left}) nonzero entries of $A$ are all equal to 1 (\emph{right}) nonzero entries of $A$ are random uniform samples on the interval $[0,1]$}
\end{figure}

We illustrate the breakdown in the sparse block-encoding methods on nonnegative sparse matrices in Figure 14. Here, we attempt to block-encode an 8-qubit matrix of nonnegative data (positive values greater than 0.7) with $s\approx 15.06$. While the sign of the approximation errors in the FABLE circuit appear to randomly vary by pixel, the same cannot be said for S-FABLE where the zero entries are all underestimated and the nonzero entries overestimated, or for LS-FABLE where all entries in the block are underestimated (with seemingly no special weighting of errors at the sites of the nonzero entries). However, if the nonnegative matrix is entry-wise multiplied by random values uniformly sampled from $[-1,1]$ as in Figure 15, we recover the improved behavior of S-FABLE and LS-FABLE. Notice also in these cases that the error plots exhibit a much more spatially agnostic distribution of positive and negative errors.

\begin{figure}
\centering
\begin{tabular}{c|c|c|c|}
        ~ & FABLE & S-FABLE & LS-FABLE \\
      \hline
      \rotatebox{90}{Block-Encoding} & \includegraphics[scale=0.5]{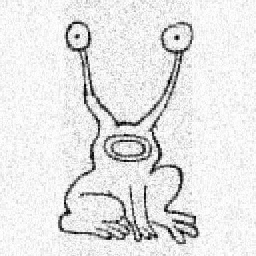} &
      \includegraphics[scale=0.5]{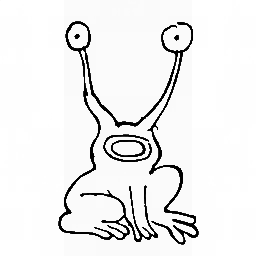} & \includegraphics[scale=0.5]{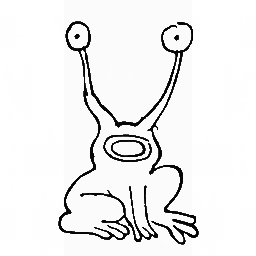} \\
      \hline
      \rotatebox{90}{Error} & \includegraphics[scale=0.5]{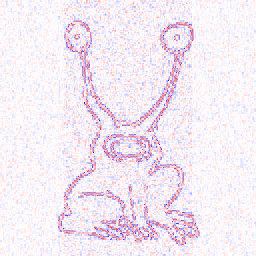} &
      \includegraphics[scale=0.5]{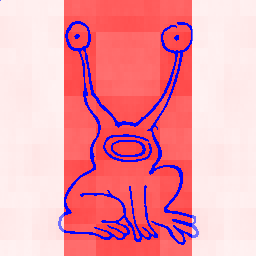} & \includegraphics[scale=0.5]{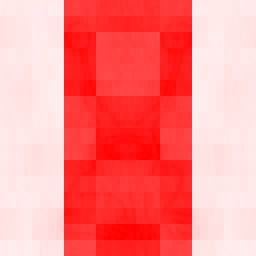} \\
      \hline
\end{tabular}
\caption{Comparison of (\emph{Left}) FABLE, (\emph{Center}) S-FABLE, and (\emph{right}) LS-FABLE for block encoding a sparse positive matrix with $n=8$ and $s\approx 15.06$. $\epsilon _F(A)=0.4562$, $\epsilon _S(A)=0.1727$, and $\epsilon _{LS}(A)=0.2022$. (\emph{Top}) Visual depiction of block-encoding approximations (\emph{Bottom}) Relative plot of errors. Red corresponds to underestimation and blue corresponds to overestimation.}
\end{figure}
        
\begin{figure}
\centering
\begin{tabular}{c|c|c|c|}
        ~ & FABLE & S-FABLE & LS-FABLE \\
      \hline
      \rotatebox{90}{Block-Encoding} & \includegraphics[scale=0.5]{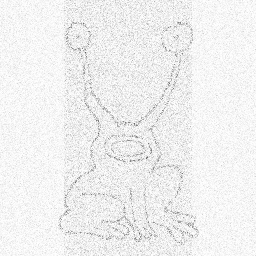} &
      \includegraphics[scale=0.5]{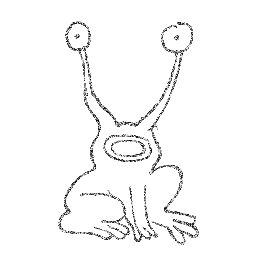} & \includegraphics[scale=0.5]{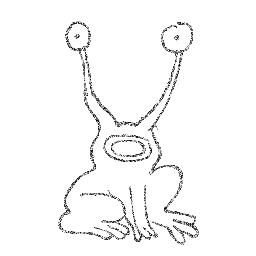} \\
      \hline
      \rotatebox{90}{Error} & \includegraphics[scale=0.5]{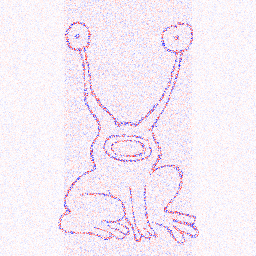} &
      \includegraphics[scale=0.5]{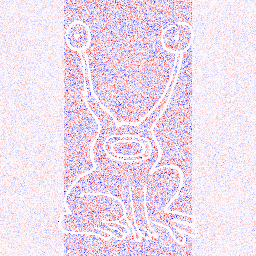} & \includegraphics[scale=0.5]{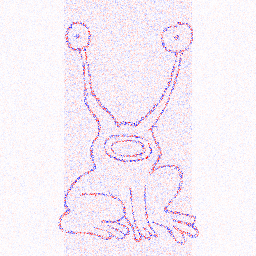} \\
      \hline
\end{tabular}
\caption{Comparison of (\emph{Left}) FABLE, (\emph{Center}) S-FABLE, and (\emph{right}) LS-FABLE for block encoding a sparse matrix (matrix from Figure 14 with each nonzero entry multiplied by a random quantity from 
$[-1,1]$) with $n=8$ and $s\approx 15.06$. $\epsilon _F(A)=4.5337$, $\epsilon _S(A)=0.0452$, and $\epsilon _{LS}(A)=0.0723$. (\emph{Top}) Visual depiction of block-encoding approximations (\emph{Bottom}) Relative plot of errors. Red corresponds to underestimation and blue corresponds to overestimation.}
\end{figure}

\section{Conclusion}\label{conclusion}

In this paper we developed two modifications of FABLE \cite{fable22} which achieve higher compression when block-encoding sparse matrices. S-FABLE uses a FABLE circuit to block-encode $HAH$ and then recovers the sparse matrix $A$ by conjugating the circuit extra Hadamard gates. Meanwhile, LS-FABLE altogether abandons the $\mathcal{O}(N^2\log{N})$ classical computation necessary to compute the angles needed in FABLE and S-FABLE by instead directly encoding a scaled copy of the sparse matrix as the rotation-gate angles themselves.

Both S-FABLE and LS-FABLE outperform FABLE on random unstructured sparse matrices. This improvement is exponential; the $L^2$ norm error of the approximation scales as $\epsilon _{LS}(A)\approx \mathcal{O}(1/N)$ using only $\mathcal{O}(N)$ rotation gates as the qubit count increases. However, this improvement breaks down as the matrices become symmetric or only take on positive values. We posit that in these highly structured cases, other more efficient block-encoding methods will likely be at our disposal.

We have shown that FABLE can be modified to accommodate highly compressible circuits to block-encode a certain subset of matrices, namely sparse matrices. There may be other common classes of matrices that do not compress well under a vanilla implementation of FABLE but could exhibit circuit contraction if the data is properly transformed. The methods developed here could also see improvement, for instance LS-FABLE uses essentially a first-order approximation to derive the rotation gate angles; it may be interesting to explore methods to access higher-order approximations still using $\mathcal{O}(N)$ classical resources. Also, this analysis only applies to real matrices; further investigation is required to consider block-encodings of complex-valued sparse matrices.

\section*{Acknowledgement}

The authors would like to thank Kevin Obenland, Peter Johnson, and Athena Caesura for their helpful comments.

\bibliographystyle{unsrt}
\bibliography{main}

\end{document}